\documentclass[a4paper,onecolumn,11pt]{quantumarticle}
\pdfoutput=1
\usepackage[utf8]{inputenc}
\usepackage[english]{babel}
\usepackage[T1]{fontenc}
\usepackage{amsmath}
\usepackage{hyperref}
\usepackage{bbold}
\usepackage[numbers,sort&compress]{natbib}
\usepackage{algpseudocode}
\usepackage{caption}
\usepackage{algorithm}
\usepackage{subcaption}

\usepackage{tikz}
\usepackage{qcircuit}
\usepackage{lipsum}
\usepackage{physics}

\begin{document}

\title{A discrete Fourier transform based quantum circuit for modular multiplication in Shor's algorithm}

\author{Abu Musa Patoary}
\affiliation{Joint Quantum Institute and Department of Physics,
University of Maryland, College Park, MD 20742, USA}
\author{Amit Vikram}
\affiliation{Joint Quantum Institute and Department of Physics,
University of Maryland, College Park, MD 20742, USA}
\affiliation{Center for Theory of Quantum Matter, Department of Physics, University of Colorado, Boulder CO 80309, USA}
\author{Victor Galitski}
\affiliation{Joint Quantum Institute and Department of Physics,
University of Maryland, College Park, MD 20742, USA}
\maketitle

\begin{abstract} 
Shor's algorithm for the prime factorization of numbers provides an exponential speedup over the best known classical algorithms. However, nontrivial practical applications have remained out of reach due to experimental limitations. The bottleneck of the experimental realization of the algorithm is the modular exponentiation operation. In this paper, based on a relation between the modular multiplication operator and generalizations of discrete Fourier transforms, we propose a quantum circuit for modular exponentiation. A distinctive feature of our proposal is that our circuit can be entirely implemented in terms of the standard quantum circuit for the discrete Fourier transform and its variants. The gate-complexity of our proposal is $O(L^3)$ where $L$ is the number of bits required to store the number being factorized. It is possible that such a proposal may provide easier avenues for near-term generic implementations of Shor's algorithm, in contrast to existing realizations which have often explicitly adapted the circuit to the number being factorized.
\end{abstract}

Shor's algorithm~\cite{shor1994algorithms} for the prime factorization of numbers provides an exponential speedup over the best known classical algorithm, the general number field sieve~\cite{lenstra1993development}. For a number with $L$ bits, the complexity of Shor's algorithm is polynomial in $L$ whereas the complexity of the general number field sieve is $O(e^{L^{1/3}(\log L)^{2/3}})$ \cite{pomerance1996tale}. As one of the few algorithms believed to show quantum advantage, Shor's algorithm has particularly received attention for its potential to break RSA encryption, which relies on the perceived difficulty of factorization~\cite{RivestRSA}. However, nontrivial practical applications have remained out of reach due to experimental limitations: for example, the largest number successfully factorized in an experimental implementation of Shor's algorithm is $21$ \cite{martin2012experimental}, which can be stored in $5$ qubits, whereas the representation of numbers used in RSA typically require around $2048-4096$ classical bits~\cite{barker2007nist} (and therefore, qubits).

To eventually achieve a practical advantage of Shor's algorithm, one would require: \romannumeral 1) the successful processing of more qubits than currently available, \romannumeral 2) better error correcting codes, and \romannumeral 3) improved efficiency in all the modules of the algorithm. The primary bottleneck for the latter is the modular exponentiation component of the algorithm, for which the most straightforward implementations require $O(L^3\log L)$ number of elementary gates for $L$ qubits~\cite{beauregard2003circuitshorsalgorithmusing, tan2024efficient}. While multiplication algorithms with better complexity exist~\cite{schonhage1971fast}, it is not clear how they may be efficiently implemented in practice. When it comes to such implementations, the circuits used in experiments conducted so far~\cite{martin2012experimental,Politi2009,Smolin2013OversimplifyingQF,PhysRevLett.99.250504,PhysRevLett.99.250505} are not generic, as the experimental implementation of modular multiplication is often simplified based on a classical foreknowledge of the factors the algorithm is meant to compute.

The challenging nature of experimental implementations has led to some alternative proposals for factorization that aim to be easier to implement for present-day demonstrations~\cite{anschuetz2018variationalquantumfactoring,yan2022factoringintegerssublinearresources,PhysRevLett.130.250601,10426927}, though these are often specialized and not expected to be an improvement over Shor's algorithm for large $L$. In contrast, a recent proposal by Regev~\cite{regev2024efficientquantumfactoringalgorithm} (and relevant extensions~\cite{ekeragartner_regevalgodiscretelog}) is in fact superior to Shor's algorithm for asymptotically large $L$, though its practical advantage for classically relevant values of $L$ (such as $2048$) remains unclear. Further, Kahanamoku-Meyer and Yao have proposed an algorithm for modular multiplication itself which, asymptotically, requires $O(L^{2+\epsilon})$ gates~\cite{kahanamokumeyer2024fastquantumintegermultiplication}. 

In this paper, we propose a quantum circuit with $O(L^3)$ elementary gates for the modular exponentiation operation. Though less efficient than the one proposed in ~\cite{kahanamokumeyer2024fastquantumintegermultiplication}, an interesting feature of our circuit is that it is composed only of Discrete Fourier Transforms (DFTs) and their closely related variants. It is also logarithmically more efficient than the ``straightforward'' circuits for modular multiplication based on the standard multiplication algorithm~\cite{nielsen2010quantum}. We note that the DFT operator is already an essential component of the \textit{phase estimation} protocol in Shor's algorithm~\cite{nielsen2010quantum}. Our circuit therefore provides an implementation of Shor's algorithm entirely in terms of DFTs (specifically, the Quantum Fourier Transform (QFT) circuit~\cite{Kitaev1995QuantumMA}) and their 
variants, without requiring additional specialized implementations of modular exponentiation.
Moreover, our circuit is generic because it doesn't require any specialization depending on the number to be factorized.
Given that the experimental implementation of the algorithms discussed above is not clear-cut, a circuit such as ours may provide a more feasible route towards short term \textit{generic} implementations of Shor's algorithm depending on the platform of choice.
 
The quantum part of Shor's algorithm is about finding the multiplicative order $r$ of a number $A$ which is co-prime to $N$, the number being factorized. The multiplicative order $r$ is the smallest positive natural number which satisfies
\begin{equation} \label{def:multiplicative order}
    A^r \pmod{N} = 1.
\end{equation}
If $r$ is even then the prime factors of $N$ are $\gcd(A^{r/2} \pm 1, N)$ where $\gcd$ denotes the `greatest common divisor' of two integers. Alternatively, one can consider the Bernoulli map $f(x) = Ax \pmod{N}$. Then the period of the orbit starting at $x=1$ is equal to the multiplicative order $r$. In Shor's algorithm a unitary quantization of this Bernoulli map, 
namely the quantum modular multiplication operator $U_A$, is utilized. The action of this operator $U_A$ on the computational basis states is given by
\begin{equation} \label{def:qmod mult}
    U_A \ket{m} = \ket{mA \pmod{N}},
\end{equation}
where $\ket{m \in \{0,1 \dots N-1\}}$ are computational basis states. Eq. \eqref{def:multiplicative order} implies that $U_A^r = \mathbb{1}$. Since $r$ can be as large as $N-1$, it is not efficient to find $r$ by repeated multiplication of $A$ or $U_A$. However the eigenvalues of $U_A$ are of the form $e^{2\pi i s/r}$ where $s$ is an integer. Consequently, a quantum phase estimation (QPE) algorithm~\cite{nielsen2010quantum} can be used to extract the value of the multiplicative order $r$. Shor's algorithm is an application of QPE where instead of repeatedly multiplying $U_A$, one uses modular exponentiation, which is the successive application of the gates $U_A^{2^k}$ controlled by auxiliary qubits, where $k \in \{0,1, \dots , k_{\max}\}$. It is notable that modular multiplication satisfies $U_A^{2^k} = U_{A^{2^k}}$. Consequently, an efficient strategy is to apply $k_{\max}$ different modular emultiplication operators with different values of $A$ --- which is particularly suitable for our circuit proposed below. Since $k_{\max}$, the maximum value of $k$, is typically of the order of $L$ where $L$ is the number of bits in which $N$ is stored, in this method one needs $O(L)$ controlled unitary modules. Each unitary module has the gate complexity $O(M(L))$ where $M(L)$ is the number of gates required to apply modular multiplication.
Apart from modular exponentiation, Shor's algorithm also needs to apply the quantum Fourier transform (QFT) for completing the phase estimation step. However, for currently known implementations, the cost of modular exponentiation is far greater than the $O(L^2)$ cost of the standard QFT circuit ~\cite{nielsen2010quantum,markov2015constantoptimizedquantumcircuitsmodular}, making the overall gate complexity of Shor's algorithm $O(L M(L))$.

It is worth comparing our circuit to the existing literature. There are many works on the construction of a quantum circuit for modular multiplication~\cite{Vedral_1996,Beckman_1996,beauregard2003circuitshorsalgorithmusing,haner2017factoringusing2n2qubits,Takahashi:2006csa,kahanamokumeyer2024fastquantumintegermultiplication}. 
All of these circuits, except the one in ~\cite{kahanamokumeyer2024fastquantumintegermultiplication}, use the quantum adder circuit by Draper~\cite{draper2000additionquantumcomputer} and an approximate QFT~\cite{coppersmith2002approximatefouriertransformuseful} to implement modular multiplication. These circuits require $O(M(L)) = O(L^2\log L)$ gates. However, Ref.~\cite{kahanamokumeyer2024fastquantumintegermultiplication} utilizes the Toom-Cook algorithm~\cite{knuth1997art} for fast multiplication to reduce the gate complexity of the circuit. In this approach, modular multiplication can be performed with $O(L^{1+\epsilon})$ gates in the asymptotic limit for arbitrarily small $\epsilon$ (and a comparable simplification is possible for the QFT component), which leads to the overall $O(L^{2+\epsilon})$ complexity mentioned above, for successive applications of modular multiplication. While our proposal for modular multiplication is $O(L^2)$, the fact that it can be be implemented in a similar way to the standard QFT circuit may simplify the real-world aspects of experimental implementation.

The core intuition behind our circuit is that the operator $U_A$ generates a permutation of the computational basis states. We know that the discrete Fourier transform (DFT) matrix consists of the orthonormal states obtained by applying DFT to the computational basis states. We further observe that rows or columns of the DFT matrix, $[F_N]_{nm} \propto e^{-2\pi i n m / N}$, can be permuted by 
rescaling the phases by an integer. Consequently, it is possible to write the modular multiplication operator $U_A$ as the product of DFT matrix and a modified DFT matrix with rescaled phases.
We will now illustrate these statements quantitatively. From Eq. \eqref{def:qmod mult} we find that the elements of $U_A$ in the computational basis are
\begin{equation} \label{elements of U_A}
     \bra{n} U_A \ket{m} \equiv [U_A]_{nm} = \delta_{n,Am-lN}
\end{equation}
where $\delta_{a,b}$ denotes the Kronecker delta function and $l \in \{0,1 \dots ,A-1\}$. We can use an identity of the Kronecker delta function to rewrite Eq. \eqref{elements of U_A} as
\begin{equation}
    [U_A]_{nm} = \frac{1}{N} \sum_{k=0}^{N-1}e^{\frac{2\pi i k (n-Am)}{N}} \equiv [F_N^{-1}.G_N]_{nm},
    \label{eq:ShorIsABaker_generalization}
\end{equation}
where $F_N$ and $G_N$ denote two matrices whose elements are 
\begin{align}
    [F_N]_{nm} &= \frac{1}{\sqrt{N}}e^{\frac{-2 \pi i nm}{N}},\\
    [G_N]_{nm} &= \frac{1}{\sqrt{N}}e^{\frac{-2 \pi i Anm}{N}}.
\end{align}
Note that $F_N$ is the DFT matrix and $G_N$ is a modified DFT matrix whose phases have been rescaled. In our context, this expression was motivated by generalizing certain relations between modular multiplication and quantum maps involving DFTs~\cite{lakshminarayan2007modular, shorbaker}\footnote{Specifically, given the setup of \cite{lakshminarayan2007modular, shorbaker}, the generalization is from $N$ being related to multiples of $A$ to arbitrary $N$, and proceeds from Eq.~\eqref{eq:ShorIsABaker_generalization} by splitting $G_N$ into different terms, each with a block structure resembling the permuted $A$-baker's quantum maps, which are thereby also generalized to arbitrary $N$. between modular multiplication and quantum maps involving DFTs~\cite{lakshminarayan2007modular, shorbaker}}.

We will implement $U_A$ by designing a circuit for these two matrices.
If $N$ is a power of $2$, then the circuit for the DFT requires only the Hadamard and the phase gates \cite{nielsen2010quantum}. However, in general we can not use that circuit here because $N$, the number to be factorized, is typically not a power of $2$. Instead one can use Kitaev's algorithm for applying the DFT operator over an arbitrary Abelian group \cite{Kitaev1995QuantumMA}. For our 
application, the Abelian group is the cyclic group of integers $\mathbb{Z}_N \in \{0,1 \dots, N-1\}$ together with the operation of addition modulo $N$. 
One restriction in our circuit is that it requires the coprime $A$ to be an odd number. The reason behind this will be clear when we describe the circuit in detail. One should be able to incorporate even coprimes with some modifications, but we do not discuss such modifications here. In section~\ref{outline of the circuit}, we will provide a general structure of our circuit, and in section~\ref{Breaking down the circuit into elementary gates}, we break down the circuit into elementary gates.

\section{Outline of the circuit} \label{outline of the circuit}
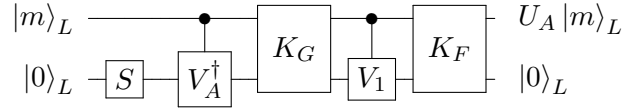
\begin{figure}[h!]
        \centering
        \[
        \Qcircuit @C=.3em @R=.7em {
        \lstick{\ket{m}_L} & \qw & \qw & \qw & \qw & \qw & \ctrl{1} & \qw &  \qw & \multigate{1}{K_G} & \qw & \ctrl{1} & \qw &  \multigate{1}{K_F} & \qw & \rstick{U_A\ket{m}_L}\\
        \lstick{\ket{0}_L} & \qw & \gate{S} & \qw & \qw &\qw & \gate{V_A^{\dagger}} & \qw  & \qw & \ghost{K_G} & \qw & \gate{V_1} & \qw &  \ghost{K_F} &  \qw & \rstick{\ket{0}_L}
        }
        \]
        \caption{The outline of the circuit for modular multiplication. There are two input registers. The first one contains the state one wants to apply modular multiplication on (state $\ket{m}_L$ in the figure) and the second one contains $L$ ancilla qubits at $\ket{0}$. The gate $S$ acts on the second register to create an equal superposition of the basis states. Then we apply two controlled phase gate $V_A^{\dagger}$ and $V_1$. After that we apply the operator $K$ which swaps the two registers and reset the second register to $\ket{0}$.}
        \label{fig: Outline of the circuit}
\end{figure}

In quantum computation, a system of multiple qubits grouped together is called a register~\cite{EkertBook}. We have two input registers each containing $L$ qubits. Define a unitary $V_A$ which
multiplies a state by a phase proportional to $A$, according to the following equation
\begin{equation} \label{Def: V_A}
     V_A \ket{m}_{1}\ket{n}_{2}=\ket{m}_{1}e^{\frac{2\pi i Amn}{N}}\ket{n}_{2}.
\end{equation}
Here $m,n \in \{0,1\dots N-1\}$ are computational basis states and the subscripts denote their register index. This is the order of registers we will stick to even if we do not explicitly mention the register number for brevity. From Eq. \eqref{Def: V_A}, we can evaluate the action of $V_A$ on an arbitrary state. Let us calculate the output of $V_A$ when the state $\ket{n}_2$ in Eq. \eqref{Def: V_A} is an equal superposition of all the basis states. We denote the equal superposition state as $\ket{\Psi_N}$ i.e.
\begin{equation} \label{equal superposition}
    \ket{\Psi_N} = \frac{1}{\sqrt{N}}\sum_{l=0}^{N-1}\ket{l}.
\end{equation}
We replace the state $\ket{n}$ in Eq. \eqref{Def: V_A} with the state $\ket{\Psi_N}$ to get
\begin{equation} \label{V_A on equal sup}
    V_A \ket{m} \ket{\Psi_N} = \ket{m} \frac{1}{\sqrt{N}}\sum_{l=0}^{N-1}e^{\frac{2\pi i Aml}{N}}\ket{l} = \ket{m} G_N^{-1} \ket{m}.
\end{equation}
We further note that 
\begin{equation} \label{V_1 on equal sup}
    V_1 \ket{m} \ket{\Psi_N} = \ket{m} F_N^{-1} \ket{m}.
\end{equation}
Now we can use Eq. \eqref{V_A on equal sup} and \eqref{V_1 on equal sup} to design the circuit for modular multiplication. The outline is given in Fig.~\ref{fig: Outline of the circuit}.

In Fig.~\ref{fig: Outline of the circuit}, the gate denoted by $S$ converts the second register at the state $\ket{0}_L$ into the equal superposition state $\ket{\Psi_N}$ i.e.
\begin{equation}
    S\ket{0}_L = \ket{\Psi_N}.
\end{equation}
After that one applies the operator $V_A^{\dagger}$ which produces the state $G_N\ket{m}$ in the second register. Then the operator $K_G$ swap the two registers and reset the second register to the state $\ket{0}$. In the next step, one applies the operator $S$ and $V_1$ to get the state $U_A\ket{m}$. Finally, similar to the previous sequence, one applies the operator $K_F$ to reset the second register to the state $\ket{0}$. This is necessary for further applications of $U_A$ with the same auxiliary register.
\section{Breaking down the circuit into elementary gates} \label{Breaking down the circuit into elementary gates}
In this section we will break down the gates in Fig. \ref{fig: Outline of the circuit} into simpler gates. First we want to perform the operation which transforms the state $\ket{0}_L$ into the equal superposition state $\ket{\Psi_N}_L$. We can create $\ket{\Psi_N}$ using a series of recursive rotations on the state $\ket{0}_L$. Let's say $N_1N_2\dots N_L$ is the binary representation of the number $N$. Then the equal superposition state can be prepared using the following algorithm \cite{Kitaev1995QuantumMA}:
\begin{algorithm}
\caption{Algorithm to create the equal superposition state $\ket{\Psi_N}$}
\label{equal superposition}
\begin{algorithmic}[1] 
    \State $\bar{N}=N$
    \For{$i=1,i++, i\leq L$}
        \If{$N_i=1$}
            \State Apply rotation $R(\theta_i)$ with $\theta_i=\tan^{-1}(\sqrt{\bar{N}/2^{L-i}-1})$ on the $i$th qubit \label{rotation angle}
            \If{$i$th ancilla is at $\ket{0}$}
                \State Apply Hadamard gate $(H)$ on $i+1$th to $L$th qubit
            \Else
               \State $i=+1$
                \State Break
            \EndIf
            \State $\bar{N} = \bar{N}- 2^{L-i}$
        \Else
            \State $i=+1$
        \EndIf
    \EndFor.
\end{algorithmic}
\end{algorithm}

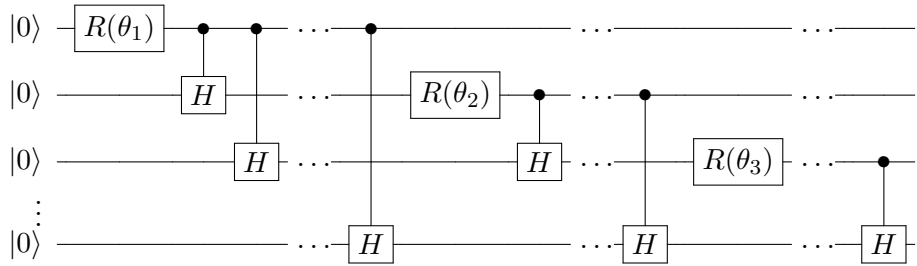
\begin{figure}[h!]
    \centering
    \[
    \Qcircuit @C=.3em @R=.7em {
    \lstick{\ket{0}} & \qw & \gate{R(\theta_1)} & \qw & \ctrl{1} & \ctrl{2} & \qw & & &\dots & & &\qw & \ctrl{4} & \qw & \qw & \qw & \qw & \qw  &&& \dots &&& \qw & \qw & \qw & \qw & \qw &&& \dots &&& \qw & \qw & \qw & \qw \\
    \lstick{\ket{0}} & \qw & \qw & \qw & \gate{H} & \qw & \qw & & &\dots & & &\qw & \qw & \qw & \gate{R(\theta_2)} & \qw & \ctrl{1} & \qw &&& \dots &&& \ctrl{3} & \qw & \qw & \qw & \qw &&& \dots &&& \qw & \qw & \qw & \qw \\
    \lstick{\ket{0}} & \qw & \qw & \qw & \qw & \gate{H} & \qw & & & \dots & & & \qw & \qw & \qw & \qw & \qw & \gate{H}  & \qw &&& \dots &&& \qw & \qw & \qw  & \gate{R(\theta_3)} & \qw &&& \dots &&& \qw & \qw & \ctrl{2} & \qw \\
    \lstick{\vdots} \\
    \lstick{\ket{0}} & \qw & \qw & \qw & \qw & \qw & \qw & & & \dots & & & \qw & \gate{H} & \qw & \qw & \qw & \qw & \qw &&& \dots &&& \gate{H} & \qw & \qw & \qw & \qw &&& \dots &&& \qw & \qw & \gate{H} & \qw
    }
    \]
    \caption{The quantum circuit to create the equal superposition state $\ket{\Psi_N}$ following algorithm~\ref{equal superposition}. One applies controlled-Hadamard $(H)$ after rotating a qubit by the angle $\theta_i$ given in line~\ref{rotation angle} of the algorithm.}
    \label{fig:circuit for superposition}
\end{figure}
From Algorithm. \ref{equal superposition} we recognize that the operator $S$ can be implemented using single qubit rotation and the Hadamard gate. The circuit to realize this algorithm is shown in Fig. \ref{fig:circuit for superposition}. The operators $V_1$ and $V_A^{\dagger}$ can be constructed from the phase shift gates $\Lambda(k)$  of the form \cite{Kitaev1995QuantumMA}
\begin{equation} \label{phase shift gate}
    \Lambda(k) = \begin{bmatrix}
                1 & 0 \\
                0 & e^{2\pi i \frac{2^k}{N}}
                \end{bmatrix},
\end{equation}
where $k \in \{0,1, \dots, L-1\}$. We break down the resetting operators $K_G$ and $K_F$ into elementary gates using the method presented by Kitaev in ~\cite{Kitaev1995QuantumMA}. The main idea is to apply inverse phase estimation for a cyclic permutation, $W = \sum \ket{m}\bra{m+1}$. Since the DFT basis states are the eigenvectors of the cyclic permutation operation, this step resets the state $\ket{m}$ to $\ket{0}$. In the following subsection we elaborate on this method. 
\subsection{Construction of the resetting operator $K_G$ and $K_F$}
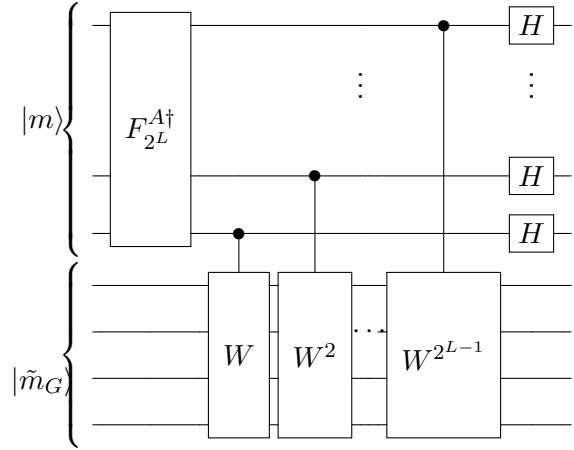
\begin{figure}[h!] 
\centering
\[
 \Qcircuit @C=.3em @R=.7em {
  \lstick{} & \qw & \multigate{4}{F_{2^L}^{A\dagger}} & \qw & \qw  & \qw   & \qw & \qw & \qw & \ctrl{5} & \gate{H} & \qw & \qw  \\
  \lstick{} & &  \pureghost{F_{2^L}^{A\dagger}}  &   &                         &         & \vdots && &  &\vdots & & \\
   \lstick{} & & \pureghost{F_{2^L}^{A\dagger}} &   &                         &         &  &&  &   & & \\
  \lstick{} & \qw & \ghost{F_{2^L}^{A\dagger}} & \qw & \qw  &\ctrl{2} & \qw                    &         \qw   & \qw & \qw & \gate{H} & \qw & \qw  \\
  \lstick{} & \qw & \ghost{F_{2^L}^{A\dagger}} & \qw & \ctrl{1}  & \qw & \qw                    &         \qw  &  \qw & \qw & \gate{H} & \qw & \qw 
  \inputgroupv{1}{5}{1.5 em}{3.3 em}{\ket{m}}\\
   \lstick{} & \qw & \qw & \qw & \multigate{3}{W}  &\multigate{3}{W^2} & \qw & \qw & \qw & \multigate{3}{W^{2^{L-1}}} & \qw & \qw & \qw \\
   \lstick{} & \qw & \qw & \qw & \ghost{W}    & \ghost{W^2} & &\cdots & &  \ghost{W^{2^{L-1}}}  & \qw & \qw & \qw        \\
  \lstick{} & \qw & \qw & \qw & \ghost{W}  & \ghost{W^2} & \qw & \qw & \qw & \ghost{W^{2^{L-1}}} & \qw & \qw & \qw    \\
  \lstick{} & \qw & \qw & \qw & \ghost{W}  & \ghost{W^2} & \qw & \qw & \qw & \ghost{W^{2^{L-1}}}  & \qw & \qw & \qw \inputgroupv{6}{9}{1.5 em}{3.3 em}{\ket{\Tilde{m}_G}}
  } 
\]

\caption{Diagram of the quantum circuit for operator $K_G$. The circuit essentially accomplishes phase estimation in reverse for the cyclic permutation $W = \sum \ket{m}\bra{m+1}$. The inputs of the circuit are the state $\ket{m}$ in the first register and the state $G_N\ket{m}$, denoted by $\ket{\Tilde{m}_G}$, in the second register The $F_{2^L}^{A}$ operator in the circuit is a DFT like transformation defined as $F_{2^L}^{A}\ket{k}= 1/\sqrt{2^L}\sum_{l=0}^{2^L-1}e^{-2\pi i Akl/2^L}\ket{l}$. Note that $F_{2^L}^{A}$ is unitary as long as $A$ is odd. The $H$ gates in the diagram refers to Hadamard gate.}
\label{KG circuit}
\end{figure}

\begin{figure}[h!] 
\centering
\[
 \Qcircuit @C=.3em @R=.7em {
  \lstick{\ket{m_1}} & \qw & \gate{H} & \gate{R_2^{A}} & \qw & &\cdots & & & \gate{R_L^{A}}& \qw & \qw & \qw & \qw & \qw & \qw & \qw & \qw & \qw & \qw & \qw & \qw \\
  \lstick{\ket{m_2}} & \qw & \qw & \ctrl{-1} & \qw \qw & \qw & \qw & \qw & \qw & \qw & \qw & \gate{H} & \gate{R_{2}^{A}} & \qw & &\cdots & & & \gate{R_{L-1}^{A}} & \qw & \qw & \qw \\
  \lstick{} & &   &   &                         &         & \vdots && &  &\vdots & & \\
   \lstick{} & &  &   &                         &         &  &&  &   & & \\
   \lstick{\ket{m_L}} & \qw & \qw & \qw & \qw & &\cdots & & & \ctrl{-4} & \qw & \qw & \qw & \qw & \qw & \qw & \qw & \qw & \ctrl{-3} & \qw & \gate{H} & \qw & \qw
  } 
\]
\caption{Circuit diagram for the operator $F_{2^L}^{A}$. In the circuit, $H$ denotes the Hadamard gate and $R_k^{A}=diag(1, e^{-2\pi i A/2^k})$ are phase gates with $k \in \{2,3\dots L\}$.}
\label{tweaked qft circuit}
\end{figure}
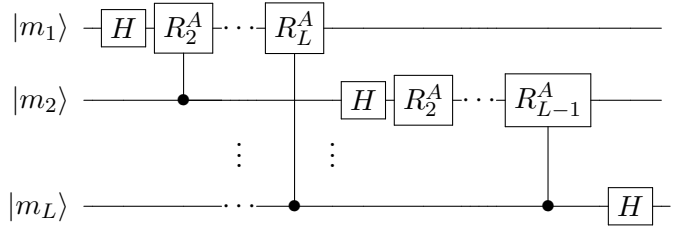

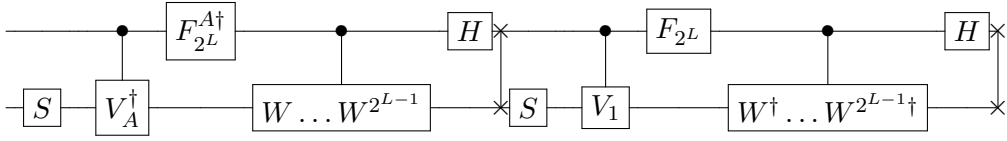
\begin{figure}[h!]
        \centering
        \[
        \Qcircuit @C=.3em @R=.7em {
        \lstick{} & \qw & \qw & \qw & \qw & \qw & \ctrl{1} & \qw &  \gate{F_{2^L}^{A\dagger}} & \qw & \ctrl{1} & \qw & \gate{H}  & \qswap & \qw & \qw & \qw &\qw & \ctrl{1} & \qw & \gate{F_{2^L}} & \qw & \ctrl{1} & \qw & \gate{H} & \qswap \\
        \lstick{} & \qw & \gate{S} & \qw & \qw &\qw & \gate{V_A^{\dagger}} & \qw  & \qw & \qw & \gate{W\dots W^{2^{L-1}}} & \qw & \qw & \qswap \qwx & \gate{S} & \qw & \qw &\qw & \gate{V_1} & \qw & \qw & \qw & \gate{W^{\dagger} \dots W^{2^{L-1}\dagger}} & \qw & \qw & \qswap \qwx
        }
        \]
        \caption{Schematic of the full circuit for modular multiplication. The operator $S$ creates an equal superposition of the basis states (see Fig.~\ref{fig:circuit for superposition} for details).$V_A$ and $V_1$ denote phase gates. $W$ is the cyclic permutation operator, $H$ is Hadmard gate and $F_{2^L}$ is QFT in $2^L$ dimensional Hilbert space. The circuit also contains SWAP gates in two different places.}
        \label{fig: expanded circuit}
\end{figure}

In this section we construct the resetting operators $K_G$ and $K_F$ using Kitaev's algorithm in ~\cite{Kitaev1995QuantumMA}. We will explicitly describe the circuit for $K_G$ only and show that $K_F$ is a special case of $K_G$. The circuit for applying the operator $K_G$ is drawn in Fig. \ref{KG circuit}. From Fig.~\ref{fig: Outline of the circuit} we see that $K_G$ is applied after applying the controlled phase gate $V_A^{\dagger}$. As a result, the inputs are the state $\ket{m}$ in the first register and the state $G_N\ket{m}$, denoted as $\ket{\Tilde{m}_G}$ for brevity, in the second register.   
Now, we apply a DFT like operator $F_{2^L}^{A\dagger}$ on the state $\ket{m}=\ket{m_1m_2\dots m_L}$. The action of  $F_{2^L}^{A}$ on the computational basis states is
\begin{equation}
    F_{2^L}^{A} \ket{k} = \frac{1}{\sqrt{2^L}}\sum_{l=0}^{2^L-1}e^{\frac{-2\pi i A kl}{2^L}}\ket{l}.
\end{equation}
We note that $F_{2^L}^{A}$ is unitary when $A$ is odd. We can realize $F_{2^L}^{A}$ using the circuit in Fig.~\ref{tweaked qft circuit}.
After applying $F_{2^L}^{A\dagger}$ on the first register we get the the state
\begin{equation}
    F_{2^L}^{A\dagger} \ket{m_1m_2\dots m_L} = \frac{1}{\sqrt{2^L}}(\ket{0}+e^{2\pi Ai 0.m_L}\ket{1})(\ket{0}+e^{2\pi Ai 0.m_{L-1}m_L}\ket{1})\dots (\ket{0}+e^{2\pi Ai 0.m_1m_2\dots m_L}\ket{1}),
\end{equation}
where $.m_j\dots m_L = \sum_{k=j}^L m_k 2^{j-k-1}$ is the binary representation of fraction. Then we apply a series of $controlled-W$ gates, denoted as $cW$. For this operation, the control and target qubits are the qubits in the first and second register respectively. After some calculation we find that
\begin{align} 
    cW (\ket{0}+e^{2\pi A i 0.m_1m_2\dots m_L}\ket{1})\ket{\Tilde{m}_G} & = (\ket{0}+e^{2\pi Ai (0.m_1m_2\dots m_L-\frac{m}{N})}\ket{1})\ket{\Tilde{m}_G} \label{Eqn for cW}\\
    cW^2 (\ket{0}+e^{2\pi Ai 0.m_2m_3\dots m_L}\ket{1})\ket{\Tilde{m}_G} & = (\ket{0}+e^{2\pi Ai (0.m_2m_3\dots m_L-\frac{2m}{N})}\ket{1})\ket{\Tilde{m}_G} \label{Eqn for cW2}\\
    \vdots & = \vdots \nonumber\\
    cW^{2^{L-1}} (\ket{0}+e^{2\pi Ai 0.m_L}\ket{1})\ket{\Tilde{m}_G} & = (\ket{0}+e^{2\pi Ai (0.m_L-\frac{2^{L-1}m}{N})}\ket{1})\ket{\Tilde{m}_G} \label{Eqn for cW2L}. 
\end{align}
We define the phases $\theta_k$ for $k \in \{1,2\dots , L\}$ as $\theta_k = 0.m_km_{k+1}\dots m_L-\frac{2^{k-1}m}{N}$. This allows us to succinctly write Eqn. \eqref{Eqn for cW} - \eqref{Eqn for cW2L} as
\begin{equation}
    cW^{2^{k-1}} (\ket{0}+e^{2\pi i 0.m_km_{k+1}\dots m_L}\ket{1})\ket{\Tilde{m}_G}  = (\ket{0}+e^{2\pi Ai \theta_k}\ket{1})\ket{\Tilde{m}_G}.
\end{equation}
It means that now the second register is in the state $\ket{\Tilde{m}_G}$ and the first register is in the product state
\begin{equation}
    \prod_{k=1}^{L}  (\ket{0}+e^{2\pi i A\theta_k}\ket{1}).
\end{equation}
Then we apply Hadamard gate on each of the qubits in the first register. It produces the state 
\begin{equation}
    \prod_{k=1}^{L} [(1+e^{2 \pi Ai \theta_k})\ket{0}+(1-e^{2 \pi Ai \theta_k})\ket{1}]\ket{\Tilde{m}_G}.
\end{equation}
Note that $\theta_k \approx 0$ which means that the two registers are approximately in the state $\ket{0}\ket{\Tilde{m}_G}$. 
Finally we swap the first and the second register to get the desired output $\ket{\Tilde{m}_G}\ket{0}$ and completes the application of $K_G$. Interestingly, we can apply the other resetting operator $K_F$ just by replacing $F_{2^L}^{A\dagger}$ with the standard quantum Fourier transform $F_{2^L}$. Now we put everything together to expand the circuit in Fig.~\ref{fig: Outline of the circuit} into Fig.~\ref{fig: expanded circuit}

 \section{Complexity}
The complexity of the circuit can be calculated from Fig. \ref{fig: expanded circuit}. We note that the realization of the $S$ gate using the circuit in Fig~\ref{fig:circuit for superposition} requires $L(L+1)/2$ elementary gates. The gates $V_1$ and $V_A$ each can be implemented using $O(L^2)$ phase-shift gates~\cite{Kitaev1995QuantumMA}. The QFT operators $F_{2^L}^{A}$ and $F_{2^L}$ each require $O(L^2)$ gates. Finally, we note that the operators $W$, $W^2$, $W^4$, $\dots$ are modular addition by a constant number. The controlled modular additions can be done in sub-quadratic time following the method in ~\cite{kahanamokumeyer2024fastquantumintegermultiplication}. Consequently, the complexity of our full circuit for modular multiplication is $O(L^2)$. In Shor's algorithm, one needs to apply upto $L$ such modular multiplication for different powers of $A$ which makes the complexity of the modular exponentiation module in Shor's algorithm $O(L^3)$.
\section{Discussion}
We have presented a quantum circuit for performing modular exponentiation, which is computationally the most expensive part of Shor's algorithm. A distinctive feature of our circuit is that it consists entirely of the QFT circuit and its variants. This is accomplished by utilizing Kitaev's algorithm for Fourier transform on the Abelian group $\mathcal{Z}_N$ with some modifications. 
Our approach is different from other proposals~\cite{kahanamokumeyer2024fastquantumintegermultiplication,Vedral_1996,Beckman_1996,beauregard2003circuitshorsalgorithmusing,haner2017factoringusing2n2qubits,Takahashi:2006csa} where the circuit is constructed by combining quantum adder circuits. For Shor's algorithm to be useful in breaking RSA encryption, the algorithm must be able to factorize large numbers $N$ that can be stored in about 2048 bits. 
The complexity of our circuit is $O(L^3)$ where $L$ is the number of bits of the number being factorized. While this is larger by $O(L^\alpha)$, where $\alpha < 1$, compared to the best known proposal for Shor's algorithm~\cite{kahanamokumeyer2024fastquantumintegermultiplication}, its advantages in reusing the structure of a single QFT implementation for modular multiplication may allow near-term experimental implementations for generic values of $N$, depending on the platform. Further, if one can adapt more efficient circuits for the QFT~\cite{EfficientQFT}, which are only logarithmically worse than $O(L)$, to the matrix $G_N$, it may be possible to realize Shor's algorithm with a complexity only logarithmically worse than $O(L^2)$.

We conclude with some general comments. It is noteworthy that the complexity of any circuit depends on the order of the coprime $A$, where smaller values are more favorable. The circuit would be significantly simplified if one can a priori choose $A$ with small order. Alternatively, it may be possible to run the circuit for a superposition of different $A$s and extract the smallest order.      

\section{Acknowledgement}
This work was supported by the National Science Foundation under Grant No. DMR-2037158, the U.S. Army Research Office under Contract No. W911NF-23-1-0241. We thank Michael Gullans for pointing out a mistake in the previous version of this paper.
\bibliographystyle{unsrtnat}
\bibliography{references}

\end{document}